\documentclass[prl, twocolumn, showpacs]{revtex4}
\usepackage{amssymb}
\usepackage{amsmath}
\usepackage{epsfig}
\usepackage{bm}

\begin{document}

%__________________________________________________________________________
%
%%\documentstyle[preprint,aps,amssymb]{revtex}  % PREPRINT format REVTEX 4.0
%\documentclass[draft,twocolumn,aps]{revtex4} % TWOCOLUMN format REVTEX 4.0
%%\documentstyle[aps]{revtex}           % GALLEY format REVTEX 4.0
%option: preprint

% this command makes pacs numbers print

%-----------------------------------------------------------------------------

%\title{\vskip -0.5in\hfill\hfil{\rm\normalsize Printed on \today}\vskip 0.4in

\title{Microwave-induced resistance oscillations as a classical memory effect}

\author{Y. M. Beltukov$^{1,2}$ and M. I. Dyakonov$^{2}$}  

\affiliation{$^{1}$Ioffe Institute, 194021, St. Petersburg, Russia\\ $^{2}$Laboratoire Charles Coulomb, Universit\'e Montpellier,
CNRS, 34095, Montpellier, France}

%\date{Received \quad}

%-----------------------------------------------------------------------------

\begin{abstract}
By numerical simulations and analytical studies, we show that the phenomenon of microwave-induced  resistance oscillations can be understood as a classical memory effect caused by recollisions of electrons with scattering centers after a cyclotron period. We develop a Drude-like approach to magneto-transport in presence of a microwave field, taking account of memory effects, and find an excellent agreement between numerical and analytical results, as well as a qualitative agreement with experiment.
\pacs{73.40.-c, 73.43.-f, 73.21.-b, 78.67.-n}

\end{abstract}

\maketitle

%-----------------------------------------------------------------------------
Nearly 20 years ago Zudov, Du, Simmons, and Reno~\cite{Zudov} and later Mani et al~\cite{Mani,Maniremark} experimentally discovered huge microwave-induced resistance oscillations (MIRO) in high-mobility two-dimensional electron gas at low temperatures and moderate magnetic fields. This spectacular phenomenon with many very unusual features has attracted a lot of interest. A detailed review of experimental results and theoretical approaches is presented by Dmitriev et al~\cite{review}.

Starting with the pioneering works~\cite{Ryzhii, Suris} which predicted oscillatory photoconductivity long before its experimental observations, the mainstream theories describe MIRO as a quantum phenomenon~\cite{review} and deal with quantum transitions between Landau levels in crossed electric and magnetic fields in the presence of electron scattering  by different types of disorder.

In this paper we demonstrate that the so-called ``displacement'' mechanism of MIRO can be understood as a {\it classical} memory effect caused by recollisions of electrons with scattering centers after one or more cyclotron periods. We propose a simple Drude-like equation taking account of such memory effects. 

The idea that memory effects due to re-collisions are important for understanding MIRO was previously put forward by Vavilov and Aleiner~\cite{Vavilov}. They derived a quantum kinetic equation including such effects and considered quantum interference of scattering amplitudes, using the self-consistent Born approximation and the Keldysh technique. Our purely {\it classical} approach is much more transparent, although based on a similar physical picture.

In strong enough magnetic fields, memory effects are known to result in classical localization when the resistivity $\rho_{xx}$ is zero~\cite{Baskin} or exponentially small~\cite{Fogler}. At {\it low} magnetic fields the magnetoresistance can be either positive~\cite{Evers} 
(soft scatterers) or negative (hard scatterers)~\cite{Jullien}. Here we consider a regime which is far from localization.

\begin{figure}
\includegraphics[width=\columnwidth]{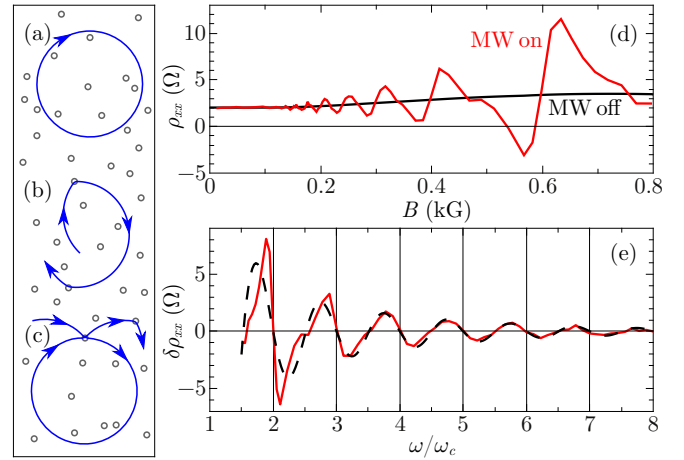}
\caption
{(Color online) Numerical simulation of classical electron magnetotransport in presence of a circularly polarized microwave radiation. Left panel: randomly distributed scattering centers and typical trajectories without (a), (b), and with re-collisions (c). Right panel: (d) numerically calculated resistivity, $\rho_{xx}$, as a function of magnetic field, with and without microwaves, (e) calculated MIRO resistivity, $\delta\rho_{xx}$, as function of the ratio $\omega/\omega_c$. Dashed line - theory, see Eq.~(17).}
\end{figure}  

We start with presenting the results of our numerical experiment, based entirely on Newton mechanics (Fig.~1), which reproduces quite well the typical experimental results for MIRO, notably the absolute negative resistance in Fig.~1(d).

We use the following input parameters. Sample size: $200\times200$~$\mu$m$^2$, impurity concentration: $N=1.1\cdot10^8$~cm$^{-2}$, Fermi energy:  $E_{\scriptscriptstyle F}= 8.6$~meV, effective mass: $m=0.067m_e$, {\it ac} field amplitude: $E_1=2$\,V/cm, {\it ac} frequency:  $\omega=2\pi\cdot 50$\,GHz, {\it dc} electric field: $E_0=0.02$\,V/cm. These parameters fairly well correspond to the typical experimental conditions.

We choose the impurity potential as: $V(r)=V_0[1-(r/r_0)^2]^{5/2}$ for $r<r_0$ and $V(r)=0$ for $r>r_0$, with $V_0=0.6E_{\scriptscriptstyle F}$ and $r_0=55$\,nm. The exact form of $V(r)$ is not really important.

Each point in Fig.~1 was obtained by averaging over $5\cdot10^8$ electron trajectories with random initial conditions. Interactions between electrons was neglected. 

The initial velocities have the zero-temperature Fermi distribution which does not noticeably change during the numerical experiment. The resistivity was defined as $\rho_{xx}=\sigma_{xx}/(\sigma_{xx}^2+\sigma_{xy}^2)$, where the conductivity tensor $\hat\sigma$ was evaluated by calculating the average electron flow caused by the {\it dc} electric field  $\bm{E}_0$. 

We have checked that the conductivity tensor calculated numerically at low magnetic field and in the absence of microwaves coincides with the predictions following from the Boltzmann equation for the chosen form of the scattering potential. The technical details of the simulation procedure will be presented elsewhere.

Electron trajectories were calculated on a PC by the velocity Verlet algorithm adapted for problems involving magnetic field~\cite{verlet} with a variable time step. A graphics processing unit (GPU) was used to increase the performance~\cite{cuda}. It takes about 2 hours to calculate each point in Fig.~1 with a  Nvidia GPU (GTX 560 model).

MIRO-like oscillations were previously obtained numerically in Ref.~\cite{Shepel} for a model involving multiple recollisions with hard disks and a special source of noise.

\begin{figure}
\includegraphics[width=\columnwidth]{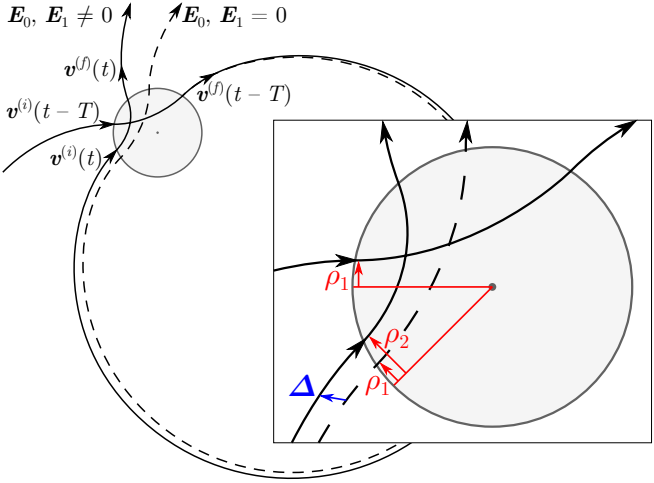}
\caption
{(Color online) Extended double collision. At time $t-T$ the electron hits the scattering center with an impact parameter $\rho_1$ and velocity $\bm{v}^{(i)}(t-T)$, which is changed to $\bm{v}^{(f)}(t-T)$ after scattering. In the absence of external fields (dashed line), after completing the cyclotron circle the impact parameter remains to be $\rho_1$, while the velocity becomes  $\bm{v}^{(i)}(t)$. The action of {\it dc} and {\it ac} electric fields during the cyclotron period produces a mismatch $\bm{\mathit{\Delta}}$, which results in changing the new impact parameter to $\rho_2$. After the second scattering, the velocity becomes $\bm{v}^{(f)}(t)$.}
\label{fig2}
\end{figure}  

In the absence of electric fields, the impact parameter and scattering angle during recollisions remans the same (Fig.~2). The crucial role of external fields is to introduce a {\it mismatch}, $\bm{\mathit{\Delta}}$, of trajectories after each cycle~\cite{Vavilov}. This mismatch consists of two parts: $\bm{\mathit{\Delta}}=\bm{\mathit{\Delta}}_0+\bm{\mathit{\Delta}}_1$ due to the actions of the {\it dc} and {\it ac} electric fields, $\bm{E}_0$ and $\bm{E}_1(t)$: 

\begin{equation}
\bm{\mathit{\Delta}}_0=2\pi\frac{e}{m}\frac{\bm{E}_0 \times \bm{\omega}_c}{\omega_c^3},\;
\bm{\mathit{\Delta}}_1(t)=\frac{e}{m}\frac{\bm{E}_1(t)-\bm{E}_1(t-T)}{\omega(\omega-\omega_c)}.  \label{eq:d1}
\end{equation}
where $e$ and $m$ are the elementary charge and effective mass, $\bm{\omega}_c=e\bm{B}/mc$, $T=2\pi/\omega_c$ is the cyclotron period. The {\it ac} field $\bm{E}_1(t)$, is assumed to be {\it circularly polarized} in the sense of cyclotron rotation~\cite{remark}.

We now introduce our Drude-like approach accounting for the memory effects related to recollisions and resulting in a simple equation for the average electron velocity.

Collisions of the type presented in Fig.~2 can be considered as {\it extended} collisions. The external fields, $\bm{E}_0$ and $\bm{E}_1(t)$, act on the electron {\it during} such extended collisions. 

Memory effects are mathematically described by equations that are non-local in time. Thus, to account for returns, the collision integral in the kinetic equation for the distribution function $f({\bm{v}},t)$ should include, terms containing this function at earlier times, $f({\bm{v}},t-nT)$,  with non-negative integer values of $n$~\cite{remark1}.

Here, we descend to the level of the Drude equation for the average electron velocity ${\bm{v}}(t)$. Within this approach, the conventional relaxation term $-{\bm{v}}(t)/\tau_{tr}$ ($\tau_{tr}$ is the transport relaxation time), describing the change of velocity at time ${\it t}$ due to collisions, must be modified to contain the velocities at previous times $t-nT$. 

Fig.~2 (drawn for $n=1$), shows that the average of velocity change during a collision at time $t$, $\delta\bm{v}=\bm{v}^{(f)}(t)-\bm{v}^{(i)}(t)$, is proportional to the average velocity at time $t-nT$.  These considerations lead to  our {\it main result}, the following Drude-like equation, accounting for memory effects caused by extended collisions: 
\begin{equation}
  \dot{\bm{v}}(t) = \bm{\omega}_c \times \bm{v}(t) - \frac{e}{m}\bm{E} - (1-p)\sum_{n=0}^\infty p^n \hat{\Gamma}^{(n)}\bm{v}(t-nT),   \label{eq:delayed-drude}   
\end{equation}
where ${\bm{E}}={\bm{E}_0}+{\bm{E}_1}(t)$ is the total electric field, the sum is over the number of recollisions $n$ (so that $n=0$ corresponds to a simple collision, and $\Gamma_{ij}^{(0)}=\delta_{ij}/\tau_{tr})$, $p$ is the probability for the electron to make a full circle unperturbed by collisions, it is also the fraction of electrons that rotate in free space and do not contribute to conductivity.

The conventional expression for the probability $p$ is~\cite{review}: $p=\exp(-2\pi/\omega_c\tau_q)$, where $\tau_q$ is the so-called quantum lifetime~\cite{remark2}. For our model, $1/\tau_q =2r_0Nv_{\scriptscriptstyle F}$. 

The tensor $\smash{\hat\Gamma^{(n)}}$ in Eq.~(2) describes the rate of velocity changes due to extended collisions with $n$ returns. It is time-dependent and generally depends on all the mismatches occurring in each of $n$ cycles. 

Equations (1, 2) describe the memory effects in the dark magnetoresistance and the {\it ac} conductivity, as well as the microwave-induced oscillations of $\rho_{xx}$ and $\rho_{xy}$. They also describe effects that are non-linear in microwave power and/or the {\it dc} field ${\bm{E}_0}$ \cite{inelastic1, inelastic2}.

{\it Dark magnetoresistance}. In the absence of the {\it ac} field, the linear in $\bm{E}_0$ magnetotransport is described by the stationary solution of Eq.~(2) with $\smash{\Gamma_{ij}^{(n)}} = \gamma_n\delta_{ij}$:
 \begin{equation}
 \bm{\omega}_c \times \bm{v} - \frac{e}{m}\bm{E}_0-\gamma\bm{v}=0, \quad\gamma= (1-p)\sum_{n=0}^{\infty} p^n\gamma_n
\end{equation}
Thus the magnetoresistance is given by the formula:
\begin{equation}
\rho_{xx}(B)/\rho_{xx}(0)=\gamma\tau_{tr}.     
\end{equation}

The parameters $\gamma_n$ can be readily evaluated, since in the absence of external fields the impact parameter $\rho$ remains the same during an arbitrary number of recollisions:
\begin{equation}
 \gamma_n = Nv_{\scriptscriptstyle F}\int [\cos(n\theta)-\cos((n+1)\theta)]\sigma(\theta)d\theta,           
\end{equation}
where $\sigma(\theta)$ is the differential scattering cross-section. Note that $\gamma_0=1/\tau_{tr}$.
Equations.~(4, 5) gives the $\rho_{xx}(B)$ dependence indistinguishable from the corresponding result of simulations in Fig.~1(c) (the ``MW off'' curve). For $p\ll 1$ and small angle scattering, when $\gamma_1=3\gamma_0$, Eq.~4 coincides with the corresponding result in Ref.~\cite{Vavilov} obtained by a quantum approach.

{\it Microwave-induced resistance oscillations}. For  $n\geq1$, the tensor $\hat{\Gamma}^{(n)}$ in Eq.~(2)  oscillates in time due to mismatches caused by the microwave field. To solve Eq.~(2), we look for a solution in the form  $\bm{v}(t)=\bm{v}+\bm{v}_1(t)$, where $\bm{v}$ is the constant part, and $\bm{v}_1(t)$ is the oscillating part induced by the {\it ac} field $\bm{E}_1(t)$:
\begin{equation}
\bm{v}_1(t) = \frac{e}{m}\frac{\bm{\omega}_c\times\bm{E}_1(t)}{\omega_c(\omega-\omega_c)}.
\end{equation}
Inserting this result into the last term of Eq.~(2) and averaging over the period of the microwave field, we obtain the following equation for the steady-state velocity $\bm{v}$:
\begin{equation}
\bm{\omega}_c \times \bm{v}-(\gamma+\hat{\widetilde{\Gamma}})\bm{v}-\frac{e}{m}(\bm{E}_0+\widetilde{\bm{E}}) = 0,  
\end{equation}
where the microwave-induced relaxation tensor $\widetilde{\Gamma}_{ij}$ and the effective  electric field $\widetilde{\bm{E}}$ are given by:
\begin{align}
\frac{e}{m}\widetilde{\bm{E}}&=(1-p)\sum_{n=1}^\infty p^n\langle\hat{\Gamma}^{(n)}\bm{v}_1(t-nT)\rangle,\\
\widetilde{\Gamma}_{ij}&=(1-p)\sum_{n=1}^\infty p^n\Big(\langle\Gamma_{ij}^{(n)}\rangle-\delta_{ij}\gamma_n\Big).
\end{align}
Here, the angular brackets denote averaging over the period of the {\it ac} field. Thus the action of microwave radiation during extended collisions (i) modifies the relaxation term and (ii) produces an effective {\it dc} electric field, $\widetilde{\bm{E}}$.

The relaxation tensor $\widetilde{\Gamma}_{ij}$ and effective field $\widetilde{\bm{E}}$ are both oscillating functions of the ratio $\omega/\omega_c$ and proportional to the power of microwave radiation. They also depend on the polarization of the microwave field $\bm{E}_1(t)$. For circular polarization the tensor $\widetilde{\Gamma}_{ij}$ is diagonal: ${\widetilde{\Gamma}}_{ij}= \delta_{ij}\widetilde{\gamma}$.

The number of terms that substantially contribute to the sums in Eqs.~(2, 8, 9) depends on the value of the probability $p$. We will assume that $p\ll 1$. Consequently, in the following we will take into account single recollisions only ($n=1$)~\cite{remark3}.

The general form of the tensor $\Gamma_{ij}^{(1)}$, depending on the vector $\bm{\mathit{\Delta}}$, is: 
\begin{equation}
\Gamma_{ij}^{(1)} - \gamma_1\delta_{ij} = \alpha\Delta^2\delta_{ij} + \beta\Delta_i\Delta_j,
\end{equation} 
where $\alpha$ and $\beta$ are functions of $\Delta^2$, $\bm{\mathit{\Delta}}=\bm{\mathit{\Delta}}_0+\bm{\mathit{\Delta}}_1(t)$ is given by Eq.~(1). To the lowest order in $\Delta$ the coefficients $\alpha$ and $\beta$ are constants that will be calculated below.

The field $\widetilde{\bm{E}}$ being  proportional to the $\it dc$ electric field ${\bm{E}}_0$, its components can be generally presented as:
\begin{equation}
    \widetilde{\bm{E}}= \varkappa_{\|} \bm{E}_0+\varkappa_{\perp}\frac{\bm{E}_0\times\bm{\omega}_c}{\omega_c},
\end{equation}
We solve Eq.~(7) with ${\widetilde{\Gamma}_{ij}}= \delta_{ij}\widetilde{\gamma}$ to find the corrections $\delta\rho_{xx}$ and $\delta\rho_{xy}$ to the longitudinal and Hall resistances respectively. Keeping only terms that are linear in $\varkappa_{\|}$, $\varkappa_{\perp}$, and $\widetilde{\gamma}$, we obtain:
\begin{equation} 
 \delta\rho_{xx}/\rho_{xx}^{(0)} = (\widetilde{\gamma}+\omega_c\varkappa_{\perp})\tau_{tr}, \;\;\;\; \delta\rho_{xy}/\rho_{xy}^{(0)}= -\varkappa_{\|},       
\end{equation}
where $\rho_{xx}^{(0)}$ and $\rho_{xy}^{(0)}$ are the conventional components of the resistivity tensor in the absence of microwaves.

While the corrections  $\delta\rho_{xx}$ and  $\delta\rho_{xy}$ are of the same order of magnitude, the microwave-induced correction $\delta\rho_{xx}$ might be comparable to, or even greater than $\rho_{xx}^{(0)}$. The correction to the Hall resistance is always relatively small.

With the help of Eqs.~(8--11) we can now determine the coefficients $\varkappa_{\perp}$, $\varkappa_{\|}$, $\widetilde{\gamma}$,  which define the microwave-induced corrections to $\rho_{xx}^{(0)}$ and $\rho_{xy}^{(0)}$ according to Eq.~(12):
\begin{align}
    \varkappa_{\perp} &= {\cal P}pr_0^2\frac{2\alpha+3\beta}{\omega_c }\frac{\pi\omega}{\omega_c}\sin\frac{2\pi \omega}{\omega_c},\\ 
    \varkappa_{\|} &= {\cal P}pr_0^2\frac{2\beta-4\alpha}{\omega_c}\frac{\pi\omega}{\omega_c}\sin^2\frac{\pi \omega}{\omega_c},\\
    \widetilde{\gamma} &= {\cal P}pr_0^2(4\alpha+2\beta)\sin^2\frac{\pi \omega}{\omega_c},
\end{align}
where ${\cal P}$ is the dimensionless microwave power:
\begin{equation}
    {\cal P} = \left(\frac{eE_1}{m}\right)^{\!2}\frac{1}{\omega^2(\omega-\omega_c)^2 r_0^2}.
\end{equation}
Finally, the microwave-induced resistivity is given by: 
\begin{gather}
    \frac{\delta\rho_{xx}}{\rho_{xx}^{(0)}} = -{\cal P}\exp\Bigl(-\frac{2\pi}{\omega_c\tau_q}\Bigr)\Big(C_1\frac{\pi\omega}{\omega_c}\sin\frac{2\pi \omega}{\omega_c}+C_2\sin^2\frac{\pi \omega}{\omega_c}\Big); \nonumber \\
    C_1=-r_0^2\tau_{tr}(2\alpha+3\beta), \quad C_2=-r_0^2\tau_{tr}(4\alpha+2\beta).
\end{gather}
Calculation of $\alpha$ and $\beta$ (see below) for the chosen form of the impurity potential $V(r)$ gives: $C_1=29.5 $, $C_2= 27.0$. 

The resulting curve for $\delta\rho_{xx}$ as a function of $\omega/\omega_c$ is presented by the dashed line in Fig.~1(e), showing a very good agreement with simulations, especially for $\omega/\omega_c>2$. The small deviations at higher magnetic field are due to the neglected terms in Eqs.~(8, 9) with $n>1$ and also to a small non-linearity in the microwave power.

Up to numerical factors which depend on the exact form of the potential $V(r)$, Eq.~(17) is similar to corresponding results in Refs.~\cite{review, Vavilov}, obtained by using quantum formalism. In Ref.~\cite{Vavilov}, where scattering by a random potential was considered, our $r_0$ in Eq.~(16) is replaced by the correlation radius $\xi$ of the scattering potential. Thus their Eq.~(6.11), like our Eq.~(17), does not contain the Planck constant $\hbar$, which is a clear indication that the  MIRO effect calculated in Ref.~\cite{Vavilov} is, in fact, {\it classical}. 

On the other hand, Eqs.~(72-74, 84) in Ref.~\cite{review} coincide with our Eq.~(17) with $C_1=C_2$ if the scatterer radius $r_0$ is replaced by the De-Broglie wavelength $\lambda$, which seems reasonable for the case when $\lambda\gg r_0$ \cite{remark4}.

\begin{figure}
\includegraphics[width=\columnwidth]{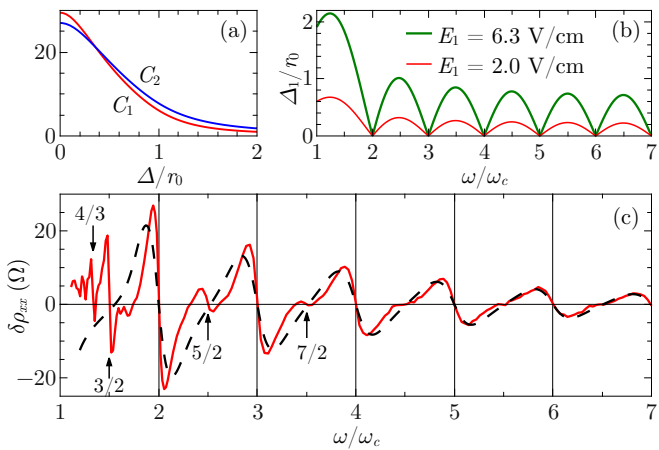}
\caption
{(Color online) MIRO for high microwave power. (a) Dependencies of $C_1$ and $C_2$ in Eq.~(17) on $\Delta/r_0$, showing the role of increasing microwave power, (b) dependence of $\Delta_1/r_0$ on the frequency ratio for $E_1=2$ V/cm (corresponding to Fig.~1) and to $E_1=6.3$ V/cm, (c) Numerically simulated MIRO for $E_1=6.3$ V/cm, dashed line - calculation using Eq.~(17) and the results in (a) and (b).} 
\label{fig3}
\end{figure}

{\it Evaluation of the parameters of an extended collision}. We briefly outline the way to determine the parameters $\alpha$ and $\beta$ in Eq.~(10). 

During the first collision with an impact parameter $\rho_1$, the initial velocity $\bm{v}^{(i)}(t-T)$ rotates by an angle $\theta_1 = \theta(\rho_1)$ and becomes $\bm{v}^{(f)}(t-T)$. After completing the cyclotron circle, the electron hits the scatterer for the second time with the velocity $\bm{v}^{(i)}(t)=\bm{v}^{(f)}(t-T)$.  

Because of the mismatch $\bm{\mathit{\Delta}}$, the new impact parameter, $\rho_2$, will differ from $\rho_1$ by the projection of the vector  $\bm{\mathit{\Delta}}$ on the direction perpendicular to $\bm{v}^{(i)}(t)$: $\rho_2=\rho_1+{\it \Delta}_\rho$.

During the second collision, the velocity rotates by the angle $\theta_2 = \theta(\rho_1+{\it \Delta}_\rho)$ and becomes $\bm{v}^{(f)}(t)$. The velocity change $\delta\bm{v}(t)=\bm{v}^{(f)}(t)-\bm{v}^{(i)}(t)$ of each electron depends on its initial impact parameter $\rho_1$ and velocity $\bm{v}^{(i)}(t-T)$. Considering ${\it \Delta}$ to be small, we expand $\delta\bm{v}$ to the second order in ${\it \Delta}_\rho$. Finally, we integrate $\delta\bm{v}(t)$ over the initial impact parameter $\rho_1$ and take the average over the distribution of the initial electron velocities, which is characterized by the average initial velocity $\bm {v}(t-T)$.

This procedure can be done both analytically and numerically, by simulating a single extended collision with $n=1$. Analytically, this results in Eq.~(10) where
$\gamma_1$ is given by Eq.~(5), the parameters $\alpha$ and $\beta$ are given by:
\begin{align}
    \alpha &= -Nv_{\scriptscriptstyle F}\int\frac{1-4\sin^2\theta}{8}(\theta'(\rho))^2d\rho, \\
    \beta &= -Nv_{\scriptscriptstyle F}\int\frac{1}{4}(\theta'(\rho))^2d\rho.
\end{align}
In the case of small angle scattering when $\theta\ll1$, we have $\beta=2\alpha$ and $C_1=C_2$. Also, $\gamma_1=3\gamma_0$.

Numerical simulation allows the calculation of $\alpha$ and $\beta$ in the general case of arbitrary $\Delta$, when the coefficients $\alpha$ and $\beta$ in Eq.~(10) become functions of $\Delta^2$.  Fig.~3(a) presents the results for the coefficients $C_1$ and $C_2$ in Eq.~(17) as functions of $\Delta/r_0$. The physical reason for the reduction of the contribution of recollisions with $n=1$ is that for large microwave power, when ${\it\Delta}/r_0 \gtrsim 1$, the electron can miss the second impact with the impurity.

{\it Nonlinear effects}. We extend our numerical simulations to study MIRO at elevated microwave power. 

The results in Fig.~3(c) (obtained for a microwave power 10 times greater than that in Fig.~1), qualitatively reproduce the main features observed experimentally, see e.\,g. Ref.~\cite{Zudov2011}. At high magnetic field, Fig.~3(c) shows oscillations at fractional values of $\omega/\omega_c$, also observed experimentally~\cite{fractions1,fractions2}.

Since we are considering effects that are linear in the $\it dc$ electric field, $|\Delta_0| \ll|\Delta_1|$ and $\Delta^2 \approx \Delta_{1}^2$. As seen from Eq.~(1), $\Delta_{1}^2$ is time-independent (this property exists for circular polarization only) and is an oscillating function of $\omega/\omega_c$, equal to zero for integer values of this ratio (Fig.~3(b)). 

Thus, the $n=1$ contribution to MIRO is suppressed between integer values of $\omega/\omega_c$, which pushes the extrema to integer values. For $\omega/\omega_c\gtrsim2$,  we obtain a good agreement between numerical experiment and the prediction of Eq.~(17), shown by the dashed line in Fig.~3(c), if the values of the coefficients $C_1$ and $C_2$ are taken from Fig.~3(a, b).

{\it In summary}, we have demonstrated that MIRO and related phenomena can be very well understood as {\it classical memory effects} caused by the action of the {\it ac} and {\it dc} fields during extended collisions, at least for some types of disorder. (This applies to the displacement mechanism. In contrast, the ``inelastic'' mechanism~\cite{review}, not considered here, strongly relies on Landau quantization and thus is truly quantum.) We have proposed a classical Drude-like equation, Eq.~(2), in which the relaxation term is modified to take account of an arbitrary number of recollisions. To our knowledge, such an approach has never been used previously. 

We have verified that the analytical results on MIRO, obtained by solving Eq.~(2), perfectly agree with the results of corresponding numerical Newton dynamics simulations (and also qualitatively agree with experiment). 

It turns out that extended collisions in the presence of external {\it dc} and {\it ac} electric fields are characterized not only by the transport cross-section, but also by additional parameters (our $\alpha$ and $\beta$) that cannot be expressed through the differential cross-section. 

Apart from minor differences, most of our physical results were previously obtained in many papers devoted to the displacement mechanism by laborious quantum calculations employing advanced theoretical techniques~\cite{review, Vavilov}. It appears, that such theories, in fact, translate into quantum language the classical physics contained in Eq.~(2). Indeed, in many cases the final results do not contain the Planck constant $\hbar$ \cite{remark5}. 

The situation is reminiscent of the conventional Drude approach to magnetotransport, which works quite well unless truly quantum phenomena, like e.\,g. Shubnikov-de Haas oscillations or weak localization, are involved. However, the only parameter in the Drude equation, $\tau_{tr}$ is expressed through the scattering cross-section, the calculation of which may, or may not, require Quantum Mechanics, depending on the relation between the de-Broglie wavelength $\lambda$ and  scatterer radius $r_0$. 

Similarly, our generalization of the Drude equation accounting for extended collisions is likely to be valid whatever is the relation between $\lambda$ and  $r_0$. We have evaluated the parameters of collision with one return using classical mechanics ($\lambda\ll r_0$). In the opposite case, the calculation of $\alpha$ and $\beta$ should be done quantum-mechanically. Since $\alpha$ and $\beta$ are not expressed through the differential cross-section, the problem of their quantum-mechanical evaluation remains open. 

In any case, the isolated problem of finding the parameters of extended collisions with a given mismatch $\bm{\mathit{\Delta}}$ is complementary to  the classical Eq.~(2). 

This work was partially supported by the Russian Foundation for Basic Research (project no. 15-02-01575).


\begin{thebibliography}{}
\bibitem{Maniremark} The correct phase of MIRO was established in Ref.~\cite{Mani}.
\bibitem{remark} Generalization for arbitrary polarization is straightforward, but will not be presented in this paper.
\bibitem{remark1} The size of impurities is assumed to be small compared to $R_c$, thus the collision time is much shorter than $T$.
\bibitem{remark2} The time $\tau_q$ defines the broadening of Landau levels, although there might be nothing quantum about $\tau_q$ itself. If the De-Broglie wavelength is smaller than the radius of the scattering center, $\tau_q$ is defined by the total classical scattering cross-section, 2$r_0$ in our case. For the chosen value of the Fermi energy and the form of $V(r)$, we have $\tau_{tr}/\tau_q =13.3$
\bibitem{inelastic1} However, we note that Eq.~(2) completely disregards inelastic processes and heating of the electrons, which were considered in Ref.~\cite{inelastic2}. The role of such processes whithin our classical approach to MIRO needs further studies.
\bibitem{remark3} The general case of an arbitrary number of re-collisions can be also analyzed, this leads to rather cumbersome formulas.  The omitted terms with $n\geq 2$ lead to resistance oscillations with higher harmonics of the ratio $\omega/\omega_c$.
\bibitem{remark4} Like similar formulas derived previously, Eq.~(17) is not applicable in the vicinity of the cyclotron resonance, see the discussion in Ref.~\cite{review}.
\bibitem{remark5} In view of our results, the term ``displacement mechanism'' based on the image of an electron shifting in space in the process of photon absorption, seems unfortunate. A more appropriate label would be the ``memory-effect mechanism''.
\end{thebibliography}
\end{document}